\documentclass[prl,reprint,amssymb,superscriptaddress]{revtex4-2}

\usepackage{xcolor}

\usepackage{bm}
\usepackage[T1]{fontenc}
\usepackage[utf8]{inputenc}
\usepackage{amsmath}  
\usepackage{amsfonts}
\usepackage{graphicx}
\usepackage{hyperref}
\usepackage{comment}
\usepackage{soul}
\hypersetup{
	colorlinks=true,
	citecolor=blue,
	linkcolor=red,
	urlcolor = blue
}

\begin{document}
	
\title{Light-induced ultrafast magnetization dynamics in van der Waals antiferromagnetic CrSBr}

\author{Ali Kefayati}
\affiliation{Department of Physics and Astronomy, University of Delaware, Newark, DE 19716, USA}
\author{Yafei Ren}
\email{yfren@udel.edu}
\affiliation{Department of Physics and Astronomy, University of Delaware, Newark, DE 19716, USA}

\begin{abstract}
We investigate the ultrafast magnetization dynamics of semiconducting antiferromagnetic CrSBr using real-time time-dependent density functional theory. In zero magnetic field, laser excitation modifies only the magnetization along the easy axis, leaving transverse components unaffected. We find that below-gap, low-fluence pulses enhance the local magnetic moments via spin transfer from nonmagnetic to magnetic atoms, increasing the N\'eel vector. In contrast, high-fluence pulses drive interlayer spin transfer between magnetic atoms, producing strong demagnetization and reducing the N\'eel vector, while S and Br atoms exhibit primarily charge transfer with weak opposite contribution to the demagnetization. An applied magnetic field qualitatively alters the response, enabling both magnitude changes and ultrafast reorientation of the magnetization. We show that the resulting layer-resolved reorientation respects a twofold rotation about the $x$-axis, exciting coherent optical magnons even under this symmetry, which modulate the relative angle between neighboring layers and periodically tune electronic properties. These results reveal a microscopic pathway for coherent magnon excitation in van der Waals magnets and establish a framework for controlling their coupled spin–charge dynamics on femtosecond timescales.
\end{abstract}

\maketitle

Controlling magnetization and its dynamics is central to both the fundamental understanding of magnetic materials and their applications in spintronics and magnonics~\cite{vzutic2004spintronics, baltz2018antiferromagnetic, vsmejkal2018topological, pirro2021advances}. The intrinsic timescale of magnetization dynamics is set by the exchange energy, typically lying in the gigahertz to terahertz range~\cite{vzutic2004spintronics, baltz2018antiferromagnetic}. Exciting magnetic materials with ultrafast laser pulses provides a means to bypass this limitation, enabling magnetization control on femtosecond to picosecond timescales~\cite{kirilyuk2010ultrafast, beaurepaire1996ultrafast, bigot2009coherent, siegrist2019light, kholid2023importance}. In the past two decades, significant progress has been made in ultrafast control of the magnetic order parameter~\cite{beaurepaire1996ultrafast, bigot2009coherent, siegrist2019light, kholid2023importance}, focusing on the demagnetization processes in transition metals, such as the ultrafast demagnetization in ferromagnets like Ni~\cite{beaurepaire1996ultrafast}, switching from uncompensated antiferromagnets to transient ferromagnets in superstructures like Co/Mn multilayers\cite{Dewhurst2018, golias2021ultrafast}, and strong terahertz emission associated with the ultrafast demagnetization~\cite{Seifert2016, Wu2021}.


The discovery and study of van der Waals magnetic materials have expanded rapidly in recent years showing diverse magnetic order and rich electronic structures, such as metals and semiconductors~\cite{du2016weak, kuo2016exfoliation, lin2016ultrathin, lee2016tunneling, huang2017layer, gong2017discovery, wu2019physical, brennan2024important, ahn2024progress, song2024future}. Ultrafast lasers provide powerful means to characterize and manipulate magnetic order and dynamics~\cite{dabrowski2022all, wu2024giant, gish2024van, zhang2022all} surpassing other techniques such as electric~\cite{huang2018electrical, jiang2018electric, zhang2020gate}, magnetic~\cite{bae2022exciton, diederich2023tunable}, mechanical~\cite{zhang2022all, hu2022doping, ni2021imaging, hu2022doping} in speed. In particular, experiments have shown that ultrafast laser pulses can serve as an efficient, nonthermal method for generating coherent magnetization dynamics, which in turn dynamically and periodically modulate the electronic and optical properties of these quantum materials~\cite{diederich2025exciton, kang2020coherent, bae2022exciton, diederich2023tunable, klaproth2023origin}. While phenomenological models provide intuitive descriptions of long-lived coherent dynamics~\cite{varela2025ultrafast}, the microscopic origin of ultrafast magnetization dynamics---its initiation and its interplay with demagnetization---remains poorly understood~\cite{chen2019revealing, he2021unravelling, li2022light, he2023ultrafast, zhou2023ultrafast, guo2024laser, huo2025ultrafast, sharma2025giant}.

In this work, we present a fully \textit{ab initio} description of ultrafast magnetization dynamics in a van der Waals magnet using real-time time-dependent density functional theory (rt-TDDFT). Our study focuses on CrSBr, a prototype semiconducting antiferromagnet showing strong coupling between magnetization and electronic excitations, as well as its large-amplitude, laser-pulse-induced magnetization dynamics~\cite{ziebel2024crsbr, bae2022exciton, diederich2023tunable}. 
%
We systematically study the effect of the laser fluence, frequency, and magnetic fields. We show that, while the previous experimental and theoretical studies always showed demagnetization of the local moments in magnetic metals~\cite{Tengdin2018, Krieger2015}, our results suggest an increase in the magnetic moments when the system is excited with a low fluence laser pulse with below gap central frequency. Above-gap or high-fluence laser induces demagnetization. In the presence of magnetic fields, we identify ultrafast reorientation of magnetization vectors that excite coherent magnon dynamics in sub-picosecond, despite that the magnon frequencies are in the order of gigahertz. These results deepen our understanding of how the coherent magnons are excited in van der Waals magnets, offering a solid starting point for future study of the subsequent magnetization dynamics.

\begin{figure*}
    \centering
    \includegraphics[width = \linewidth]{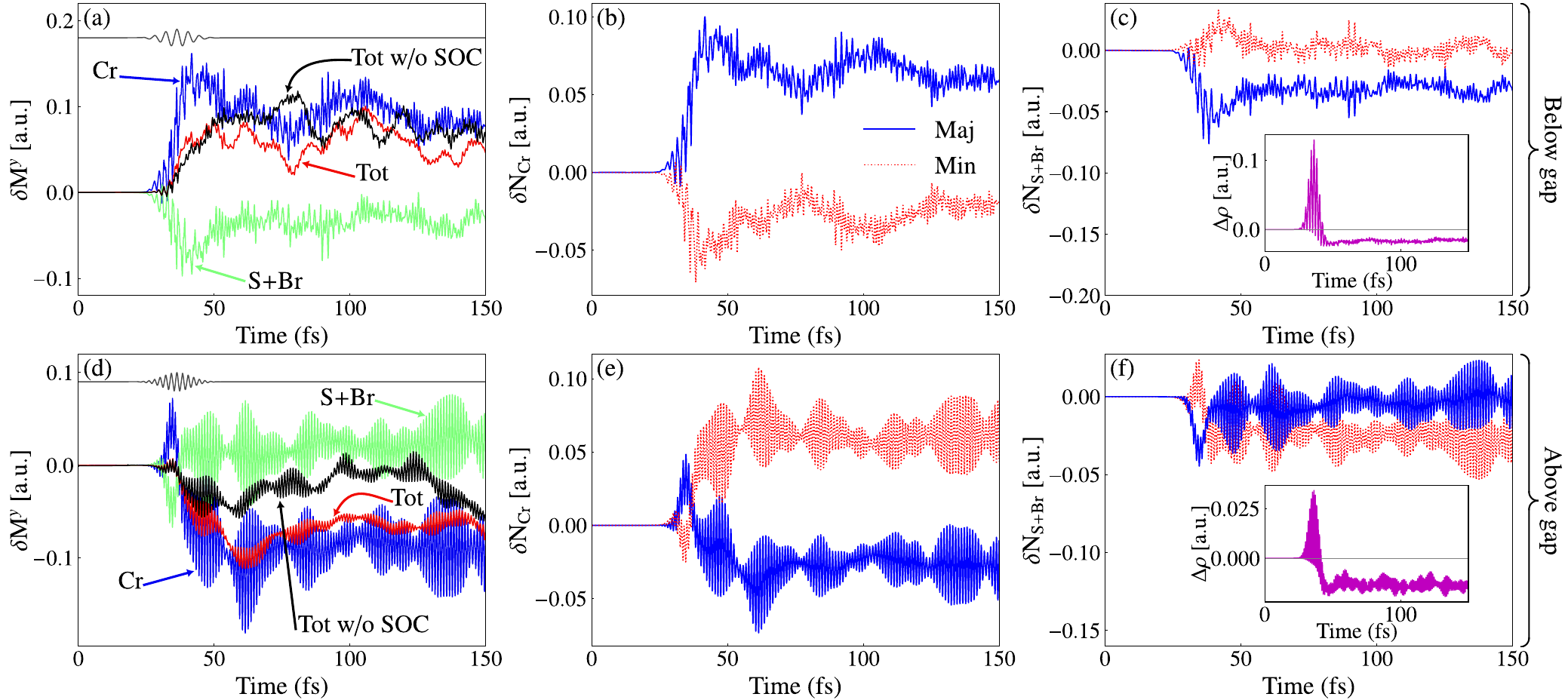}
    \caption{Magnetization dynamics in a magnetic layer (L1) in laser-excited CrSBr at low fluence ($F=5~$mJ/cm$^2$). (a)-(c) are for below-gap laser frequency. (d)-(f) are for above-gap laser frequency. (a) and (d): Time evolution of magnetic moment along $y$ direction in L1. The total magnetic moments with and without SOC are shown in red and black, separately. (b) and (e): Spin-resolved charge density in Cr atoms. (c) and (f): Spin-resolved charge density in S and Br atoms. Insets in (c) and (f): Excited free carriers densities in the interstitial region. Maj: majority spin. Min: minority spin.}
    \label{fig:fig1}
\end{figure*}

\textbf{\textit{Formalism.---}} %
We employ the start-of-the-art rt-TDDFT method to study the ultrafast magnetization dynamics excited by a laser pulse. The effect of laser pulse is captured by the time-dependent Kohn-Sham equation for Kohn-Sham orbitals $\psi_j(\bm{r},t)$ as Pauli spinors  (using $\hbar=1$)
\begin{eqnarray}\label{eq:TDDFT}
    i\frac{\partial \psi_j}{\partial t} & = & \bigg [ \frac{1}{2} \big(\bm{p}+\frac{\bm{A}(t)}{c} \big)^2 + v_s(\bm{r},t) + \frac{1}{2c} {\bm \sigma} \cdot \bm{B}_s(\bm{r},t)\nonumber\\
    & + &\frac{1}{4c^2}{\bm \sigma} \cdot \big( \nabla v_s(\bm{r},t)\times \bm{p}\big) \bigg]\psi_j(\bm{r},t),
\end{eqnarray}
where $\bm{p}=-i\nabla$ is the momentum operator, $\bm{A}(t)$ is the vector potential of the laser pulse, and $c$ is the speed of light. $v_s(\bm{r},t) = v_{\mathrm{ext}}(\bm{r},t) + v_{\mathrm{H}}(\bm{r},t) + v_{\mathrm{XC}}(\bm{r},t)$ is the effective time-dependent Kohn-Sham potential contributed by external potential from nuclei $v_{\mathrm{ext}}$, Hartree potential $v_{\mathrm{H}}$, and exchange potential $v_\mathrm{XC}$. $\bm{B}_s(\bm{r},t)=\bm{B}_\mathrm{ext}(t)+\bm{B}_\mathrm{XC}(t)$ is the time-dependent Kohn-Sham magnetic field with $\bm{B}_\mathrm{ext}$ the external static magnetic field and $\bm{B}_\mathrm{XC}$ the exchange magnetic field. 
\mbox{${\bm \sigma}=(\sigma_x, \sigma_y, \sigma_z)$} are Pauli matrices. The last term describes spin-orbit coupling. 

These Kohn-Sham orbitals give rise to particle density  \mbox{$n(\bm{r},t)=\sum_j \psi_j^\dagger(\bm{r},t) \psi_j(\bm{r},t)$} and spin magnetization \mbox{$\bm{m}(\bm{r},t)=\sum_j \psi_j^\dagger(\bm{r},t) {\bm \sigma} \psi_j(\bm{r},t)$}, so that the magnetic moment per unit cell is given by \mbox{$\bm{M}(t)=\int\!\! d^3 r\, \bm{m}(\bm{r},t)$}, integrated over the unit cell. 
Microscopic magnetization dynamics from purely electronic mechanisms is given by~\cite{elliott2020microscopic}
\begin{equation}
   \begin{split}
    \partial_t M_y(t) = \frac{1}{2c^2} \int d^3\bm{r} \big[ {\nabla} v_s \times \bm{j}_x \big]_z - \big[ {\nabla} v_s \times \bm{j}_z  \big]_x
   \end{split} 
   \label{Eq1}
\end{equation}
where $\bm{j}_{\alpha}(\bm{r},t)= \sum_i \psi_i^{\dagger}{\sigma}_\alpha \bm{j} \psi_i$ is the spin current with spin polarized along the $\alpha$-direction with $\bm{j}$ the local current density operator. Equation~\eqref{Eq1} demonstrates that magnetization dynamics is driven not only by external fields, but also by spin currents from optically excited carriers as well as spin-orbit coupling that is manifested as spatial derivatives of the potential. In this study, the ions are assumed fixed in the time-dependent calculations and we only focus on about 100 fs after the optical excitation where the most relevant dynamics is driven by the electronic system and this approximation is to a good extend valid.


We employ adiabatic local spin density approximation (ALSDA) for exchange functional within full-potential augmented plane-wave method as implemented in the ELK code~\cite{Dewhurst2016,elk}. The calculation is done in two stages. First, the ground state is obtained using static noncollinear DFT calculations. CrSBr is an A-type vdW antiferromagnetic semiconductor. The intralayer magnetic order is ferromagnetic with two Cr atoms, e.g., Cr1 and Cr2 in Fig.~\ref{fig:fig3}(a), in one unit cell.
The interlayer magnetic order is antiferromagnetic where two magnetic layers with opposite spin orientations are labeled as layer 1 (L1) containing Cr1 and Cr2 and layer 2 (L2) containing Cr3 and Cr4. CrSBr has a triaxial magnetic anisotropy with easy axis along $y$-direction~\cite{yang2021triaxial, ziebel2024crsbr}.
We use DFT+U with ${U=4.0~}$eV, ${J=1.6~}$eV, and fully localized limit double counting scheme is applied. The ground-state band gap is $E_g=1.49~$eV and the $(M_{\rm Cr}^x, M_{\rm Cr}^y, M_{\rm Cr}^z)\approx(0.0,\pm2.75,0.0)\mu_B$ magnetic moment for Cr atoms in agreement with experimental studies ~\cite{klein2023bulk, lopez2022dynamic}. Throughout the manuscript, $x$, $y$, and $z$ directions are along $a$, $b$, and $c$-axis, respectively~\cite{yang2021triaxial, ziebel2024crsbr}.
Next, the time-dependent calculation is performed using the ground-state Hamiltonian of the first step with a Gaussian femto-second laser pulse. In this study, the laser pulse is polarized along $y$-direction traveling along $z$-direction with wavelength of 1240 nm (below the band gap $E_g$, $\hbar\omega=0.67E_g$) and 620 nm (above the band gap, $\hbar\omega=1.34E_g$), full width at half maximum of 12 fs, and different laser pulse energies. The simulation time is 150 fs and a time step of 0.6 attosecond is chosen. The grid of $\bm{k}$ vectors is chosen as $8 \times 6 \times 2$. 
Since the wavelength of applied laser light is much larger than the supercell, we assume dipole approximation and disregard spatial dependence of the vector potential $\bm{A}(t)$. In calculations with the external magnetic field, a static external magnetic field ($\bm{B}_\mathrm{ext}$) is added to the Hamiltonian in both ground-state and time-dependent simulations.

\textbf{\textit{Low fluence excitation.---}} 
At low fluence ($F=5~$mJ/cm$^2$), we find that the magnetization change shows a strong dependence on the laser frequency. When the laser frequency is below the band gap, we find that the local magnetic moments of the Cr atoms are enhanced. Fig.~\ref{fig:fig1}(a) displays the time evolution of the magnetic moment of the Cr atom ($\mathrm{M_{Cr}^y}$) in one of the magnetic sublattices along the $y$ direction (the easy axis). The magnetic moments of Cr atoms in the other layer are opposite leading to zero net magnetization, guaranteed by the two-fold rotation symmetry about $a$-axis. The laser pulse is illustrated by the black curve that lasts for about 12 fs. After the pulse, $\mathrm{M_{Cr}^y}$ increases by about 2\%. The blue solid line and red dotted line show the calculations with and without SOC, respectively. One can find that the effect of SOC is weak and does not change the qualitative behavior of the evolution of $\mathrm{M_{Cr}^y}$.

The increase of the magnetic moment arises from spin transfer between magnetic and nonmagnetic atoms. As shown in Figs.~\ref{fig:fig1}(b) and \ref{fig:fig1}(c), the charge density of majority spin in Cr atom increases while that in S and Br atoms in the same layer decreases. The trend is opposite for spin minority. This indicates a spin transfer from S and Br to Cr, increasing the magnetic moment of the Cr atoms. In addition, we note that the reduction of spin in intralayer S and Br atoms is smaller than the increase of spin in Cr atom, suggesting that there is interlayer spin transfer (see Figure S9 in Supplemental Material~\cite{SM}). 
This is in contrast to the spin transfer in metallic materials with both magnetic and nonmagnetic atoms where the spin transfer typically induces demagenetization where the magnetic momentum of magnetic atoms reduces~\cite{Dewhurst2018}. Here we show that in the antiferromagnetic semiconductor, the optical laser pulsed induced inter-site spin transfer can also increase the magnetic moment of magnetic atoms. This mechanism is distinct from the recent study in ferromagnetic van der Waals monolayers where the SOC is necessary for the magnetization enhancement~\cite{sharma2025giant}.

When the light frequency is above the band gap, however, demagnetization of the Cr atoms happens as shown in Fig.~\ref{fig:fig1}(d). 
Figs.~\ref{fig:fig1}(e) and \ref{fig:fig1}(f) show that the averaged atomic spin density of S and Br decreases, whereas the spin density of Cr first increases and then reduces to amounts less than the equilibrium in the majority spin channel. In addition, the intralayer spin increase of Br and S is much weaker than the spin decrease in Cr atom, indicating that the interlayer spin transfer dominates the demagnetization. 
We also computed the charge density in the interstitial region, as shown in the insets of Figs.~\ref{fig:fig1}(c) and \ref{fig:fig1}(f) for below- and above-gap frequencies, respectively. The laser pulse is found to reduce the electron number in the interstitial region, indicating a corresponding increase in local electrons, consistent with the net electron gain on Cr atoms. However, whether the spins of these electrons align parallel or antiparallel to the magnetization direction depends on the laser frequency. 


\begin{figure}
    \centering
    \includegraphics[width = \linewidth]{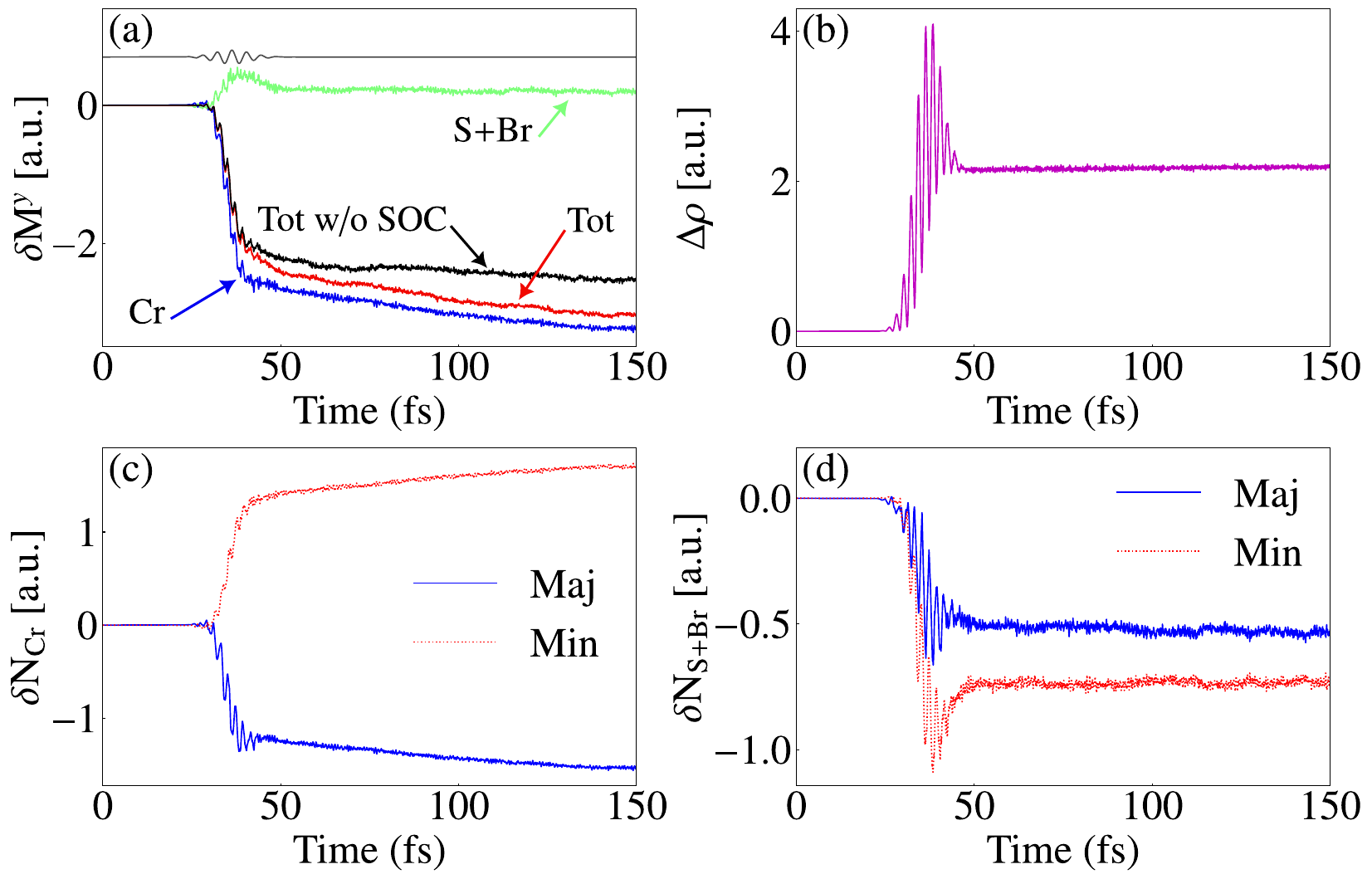}
    \caption{(a) Magnetization dynamics in one magnetic layer (L1) in the laser-excited CrSBr at high fluence ($F=100~$mJ/cm$^2$) with below gap frequency. (b) Free carrier in the interstitial region. (c) and (d): The spin-resolved charge density for Cr and S+Br atoms, respectively. Maj: majority spin. Min: minority spin. }
    \label{fig:fig2}
\end{figure}

\textbf{\textit{High fluence excitation.---}}
In the high fluence cases ($F=100~$mJ/cm$^2$), we find strong demagnetization that shows weak dependence on the laser frequency. We present the case with below-gap frequency and similar results for the above-gap case are presented in Supplemental Material~\cite{SM}. Although the laser pulse does not induce a net magnetization due to the two-fold rotation symmetry about the $a$-axis, it causes a strong reduction of the N\'eel vector. Fig.~\ref{fig:fig2}(a) shows the magnetic moment of a Cr atom in one magnetic layer, with the opposite moment in the other layer. As the laser pulse interacts with the system, the magnetic moment of Cr atoms is significantly reduced, directly contributing to the suppression of the N\'eel vector. Fig.~\ref{fig:fig2}(c) shows the spin-resolved charge transfer. The numbers of spin-majority and spin-minority electrons in Cr atoms change in a nearly symmetric fashion, indicating a spin-like transfer, i.e., net spin transfer dominates over net charge transfer.
For nonmagnetic atoms (S and Br), the net magnetic moments show a slight initial enhancement and then decay to negligible values. These atoms show a charge-like transfer. As shown in Fig.~\ref{fig:fig2}(d), in S and Br atoms, both spin-majority and spin-minority electron numbers decrease, leading to a net loss of charge with relatively smaller changes in spin. This charge-like transfer in S and Br atoms generates free carriers in the interstitial region as shown in Fig.~\ref{fig:fig2}(b).

It is noted that the demagnetization shows a weak dependence on the SOC. In the absence of SOC, the spin-flip scattering is forbidden. As the intralayer charge transfer preserves the total magnetic moment, the strong reduction of magnetic moment in one magnetic layer, about 50\%, must be attributed to the interlayer spin transfer. The presence of SOC does not show qualitative influence. The SOC starts to manifest itself near the end of the laser pulse, which introduces a weak correction of the demagnetization afterward and does not change the qualitative behavior. The demagnetization in one magnetic layer also shows weak dependence on the light frequency. We find similar qualitative behavior with above-gap frequency~\cite{SM}. Quantitatively, higher demagnetization in Cr atoms whereas magnetization increase in S and Br are observed with higher frequency.


\begin{figure}
    \centering
    \includegraphics[width = \linewidth]{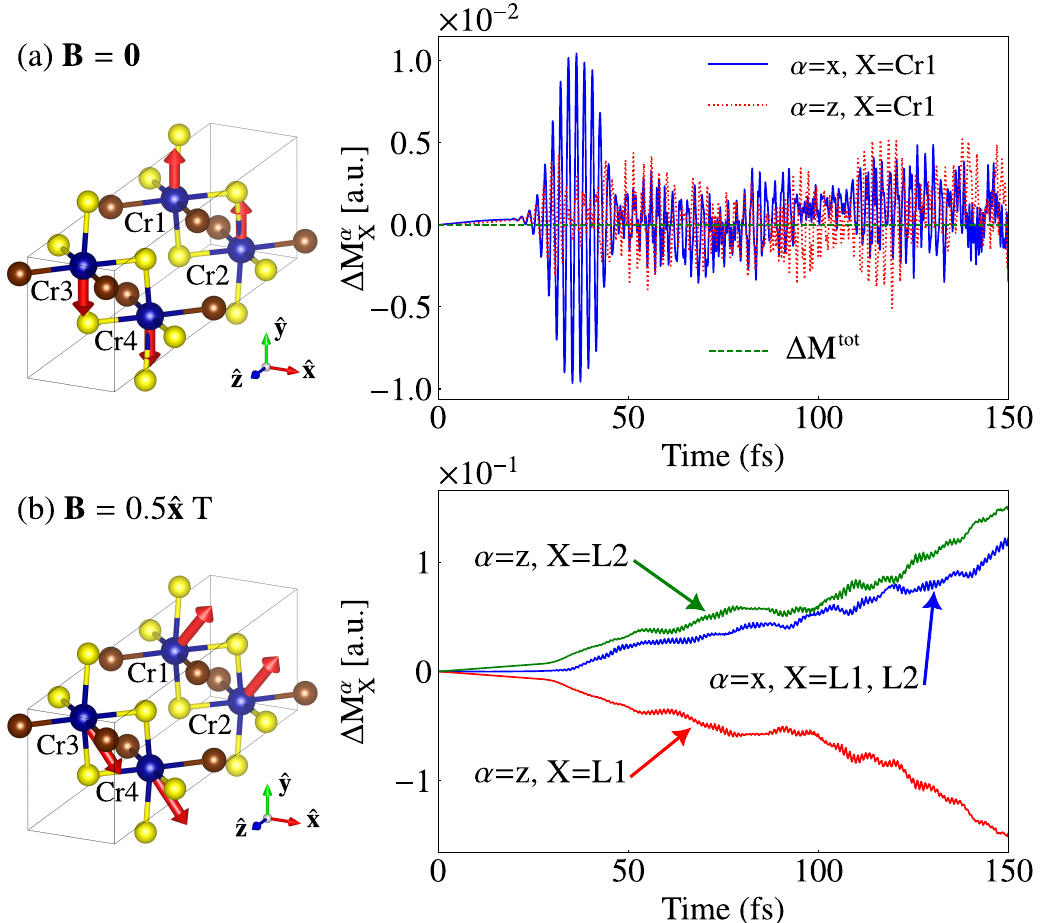}
    \caption{(a) Magnetic moment dynamics along $x$ and $z$ directions in Cr1 atom excited by a low fluence laser ($F=5~$mJ/cm$^2$) with above-gap frequency at zero magnetic field. The magnetic moment components along $x$ and $z$ in Cr2 are opposite leading to zero total values ($\Delta M^{\rm tot}$). Cr1 and Cr2 are in layer 1 (L1) with equilibrium magnetic moment pointing along $y$ direction as shown by the red arrows. Cr3 and Cr4 are in layer 2 (L2) with opposite magnetic moment. (b) Magnetic moment dynamics along $x$ and $z$ directions in L1 (X=Cr1+Cr2) and L2 (X=Cr3+Cr4) under an external magnetic field ($\bm{B}_\mathrm{ext}$). $z$-components in L1 and L2 are opposite. $x$-components in L1 and L2 are the same.}
    \label{fig:fig3}
\end{figure}

\textbf{\textit{Ultrafast magnetization reorientation with magnetic fields.---}}
The presence of a magnetic field profoundly alters the ultrafast magnetization dynamics. In the collinear spin configuration without a field, the laser pulse induces changes in the magnetization along the equilibrium direction. By contrast, an applied magnetic field enables an ultrafast reorientation of the magnetization vector, generating finite components perpendicular to the equilibrium direction. This transverse response modifies the relative spin angles on a sub-picosecond timescale, enabling an ultrafast initiation of spin dynamics and control of electronic structure and excitations coupled to the spin configuration.

We begin with the low-fluence regime. The magnetization dynamics along the easy axis with a magnetic field shows behavior similar to the zero-field case (see Fig. S7 in Supplemental Material~\cite{SM}). Below we focus on the transverse response. At zero magnetic field $\bm{B}_{\rm ext}=0$, the net transverse components in one magnetic layer are zeros. In the unit cell of one layer, there are two Cr atoms (Cr1 and Cr2). Fig.~\ref{fig:fig3}(a) shows the transverse magnetic moment of Cr1 that show fast oscillation. Nevertheless, the total magnetic moment of Cr1 and Cr2 vanishes. The presence of a magnetic field alters the equilibrium spin configuration, inducing spin canting in $x$-$y$ plane as shown in Fig.~\ref{fig:fig3}(b). The presence of a laser pulse changes the components of magnetic moment in $x$ and $z$ directions. As shown in Fig.~\ref{fig:fig3}(b), the net magnetization along $z$ direction in Cr1-Cr2 layer is opposite to that in Cr3-Cr4 layer. In contrast, the magnetic moment components along $x$ direction are the same for both layers. Such a spin configuration is invariant under a two-fold rotation operation around $x$-axis, along which the magnetic field is applied. This configuration analogies the presence of coherent magnons that are even under the operation, i.e., the so called bright magnons in CrSBr. While the magnon frequency is in the order of 30 GHz~\cite{bae2022exciton, diederich2023tunable}, the laser pulse induces a spin reorientation in 100 fs.

\begin{figure}
    \centering
    \includegraphics[width = \linewidth]{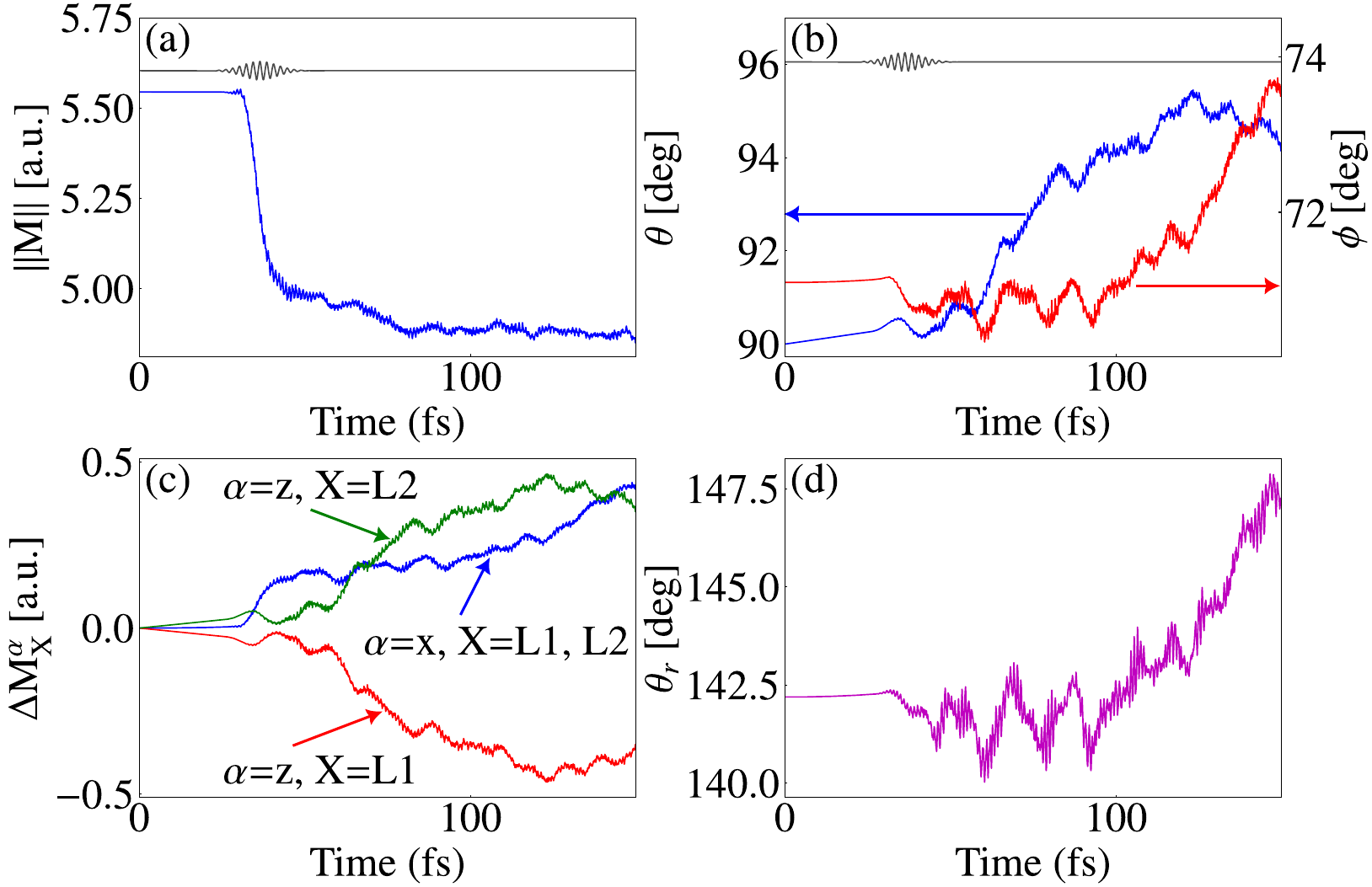}
    \caption{(a) Magnitude of the total magnetic moment in a unit cell of L1 excited by a high-fluence laser ($F=100~$mJ/cm$^2$) with above-gap frequency. (b) Azimuthal ($\theta$) and polar ($\phi$) angles of the magnetization in L1. (c) Magnetic moment dynamics along $x$ and $z$ directions in L1 and L2. (d) Relative angle between the magnetizations in two layers.
    A magnetic field $\bm{B}_\mathrm{ext}=0.5\hat{\bm{x}}~$T is applied in the calculation.}
    \label{fig:fig5}
\end{figure}

Then we turn to the high-fluence cases. The presence of a magnetic field suppresses the laser induced demagnetization while enhances the magnetization reorientation.
Fig.~\ref{fig:fig5}(a) shows the net magnetic moment amplitude in magnetic layer 1 (L1). After the laser pulse, it reduces by about $10$ to $20\%$ that is much smaller than the $50\%$ demagnetization shown in Fig.~\ref{fig:fig2}(a). This is because the demagnetization is dominant by the interlayer hopping induced spin transfer as discussed in the previous section whereas the effective interlayer hopping is suppressed by the spin canting~\cite{wilson2021interlayer}. 

Figure~\ref{fig:fig5}(b) shows that a high-fluence laser pulse can effectively change the orientation of the magnetization vector in the presence of a magnetic field. This figure shows the time-evolution of the azimuthal and polar angles of the magnetization in L1. In the absence of the laser pulse, the magnetization lies in $x$-$y$ plane with an azimuthal angle $\theta=90$. The polar angle $\phi$ indicates that the spin deviates away from the easy axis due to the magnetic field. After the laser pulse, the azimuthal angle changes by about 10 degree at within 150 fs that continuously increases. This indicates a significant change in the spin configuration in an ultrafast time scale. The magnetization in layer 2 (L2) also changes. As shown in Fig.~\ref{fig:fig5}(c), the changes of magnetization in L1 and L2 along $z$-direction are opposite while that along $x$-direction are the same. This induces an ultrafast change of the relative angle $\theta_r$ between the magnetization in L1 and L2 as shown in Fig.~\ref{fig:fig5}(d). Such a spin configuration also corresponds to the generation of large amplitude coherent magnons in this system.

The ultrafast generation of coherent magnons challenges the often-invoked analogy with coherent phonons. 
Phonons, which are quantized lattice vibrations, originate from collective displacements of heavy nuclei and are thus inherently limited by the inertia of the lattice, with dynamics constrained by their characteristic vibrational frequencies~\cite{ruhman1988coherent, weiner1991femtosecond, zeiger1992theory}. In contrast, magnons are quantized spin waves, arising from collective reorientations of electronic spins. Since spins are emergent electronic degrees of freedom, they can be reconfigured almost instantaneously through ultrafast changes in electronic states, enabling responses on femtosecond timescales that is much faster than the intrinsic dynamics of magnetization. This fundamental distinction underscores that coherent magnon excitations are not simply magnetic analogues of phonons, but rather represent a unique pathway for ultrafast spin manipulation beyond the mechanically limited dynamics of the lattice.

\textbf{\textit{Summary.---}}We studied the ultrafast dynamics of semiconducting antiferromagnetic CrSBr using the rt-TDDFT formalism. 
In the absence of a magnetic field, laser pulses only change the magnetization along the easy axis direction while the transverse components are not excited. We find that a low-fluence laser pulse with below-gap frequency can enhance the local magnetic moment and thus increase the N\'eel vector due to the spin transfer from nonmagnetic to magnetic layers. A high-fluence laser pulse, however, induces significant demagnetization in each magnetic layer and thus reduces the N\'eel vector by interlayer spin transfer between magnetic atoms. The nonmagnetic atoms, i.e., S and Br, however, show charge-like transfer that has weak contribution to demagnetization whereas generates free carriers. 

The presence of a magnetic field significantly changes the ultrafast magnetization dynamics. Beside the change of the magnetitude of the magnetization, the magnetic field enables the ultrafast reorientation of the magnetization vector. The reoriented magnetizations in neighboring magnetic layers respect the two-fold rotation about $x$-axis, corresponding to the generation of coherent magnons on the optical-magnon mode that is even under this symmetry operation. This mode changes the relative angle between neighboring magnetic layers, enabling periodic modulation of the electronic properties. 
These results deepen our understanding of how the coherent magnons are excited in van der Waals magnets and highlight the fundamental difference between coherent excitations of magnons and phonons, offering a solid starting point for future study of the subsequent magnetization dynamics.



\acknowledgments
\textbf{\textit{Acknowledgments---}} This research was supported by the US National Science Foundation (NSF) through the University of Delaware Materials Research Science and Engineering Center, DMR-2011824, and the University of
Delaware (UD) Research Foundation Strategic Initiative
Award. The supercomputing time was provided by DARWIN (Delaware Advanced Research Workforce and Innovation Network), which is supported by NSF Grant No. MRI-1919839.

%

\end{document}